\documentclass[aps,pra,twocolumn,showpacs,floatfix]{revtex4}
\usepackage{bm}
\usepackage{graphicx} % for PDF output
\usepackage{amsmath}
\begin{document}
\title{Multichannel Quantum Defect Theory for cold molecular collisions}
\author{James F. E. Croft}
\affiliation{Department of Chemistry, Durham University, South Road,
Durham, DH1~3LE, United Kingdom}
\author{Alisdair O. G. Wallis%\footnote{Present address: Chemistry
%Department, University of British Columbia, 2036 Main Mall, Vancouver,
%British Columbia, Canada V6T 1Z1}
}
\affiliation{Department of Chemistry, Durham University, South Road,
Durham, DH1~3LE, United Kingdom}
\author{Jeremy M. Hutson}
\affiliation{Department of Chemistry, Durham University, South Road,
Durham, DH1~3LE, United Kingdom}
\author{Paul S. Julienne}
\affiliation{Joint Quantum Institute, NIST and the University of
Maryland, Gaithersburg, Maryland 20899-8423, USA}

\date{\today}

\begin{abstract}
Multichannel Quantum Defect Theory (MQDT) is shown to be capable of
producing quantitatively accurate results for low-energy atom-molecule
scattering calculations. With a suitable choice of reference potential
and short-range matching distance, it is possible to define a matrix
that encapsulates the short-range collision dynamics and is only weakly
dependent on energy and magnetic field. Once this has been produced,
calculations at additional energies and fields can be performed at a
computational cost that is proportional to the number of channels $N$
and not to $N^3$. MQDT thus provides a promising method for carrying
out low-energy molecular scattering calculations on systems where full
exploration of the energy- and field-dependence is currently
impractical.
\end{abstract}

%\pacs{34.50.-s,34.50.Cx,37.10.Mn,37.10.Pq}

\maketitle

\fbox{\parbox{0.95\linewidth}{{\em Note for copy editor:} We have been
very careful to make correct use of Roman and italic subscripts and
superscripts, with Roman for abbreviations and italic for mathematical
indices. Please do not change all our subscripts and superscripts to
italic. Also note that subscript lower-case o is an abbreviation for
``open" and should not be changed to zero.}}
\section{Introduction}
\label{intro}

The creation of the first dilute atomic Bose-Einstein condensates
(BECs) in 1995 \cite{Anderson:1995, Davis:1995} led to enormous
advances in ultracold atomic physics. There is now great interest in
producing samples of cold molecules, at temperatures below 1~K
\cite{Doyle:2004, Krems:IRPC:2005, Friedrich:2009}, and ultracold
molecules, at temperatures below 1~mK \cite{Hutson:IRPC:2006,
Ni:KRb:2008, Ospelkaus:react:2010, Danzl:ground:2010}. There are many
potential applications of ultracold molecular samples, amongst which
are high-precision measurements \cite{Hudson:2002, Bethlem:2009},
quantum computation \cite{DeMille:2002} and ultracold chemistry
\cite{Bell:2009}.

Understanding atomic and molecular interactions and collisions is
essential to the study of cold and ultracold molecules. For example,
methods such as buffer-gas cooling \cite{Weinstein:CaH:1998} and Stark
deceleration \cite{Bethlem:IRPC:2003} can produce cold molecules with
temperatures between 10~mK and 1~K. However, a second-stage cooling
method is needed to bring the molecules into the ultracold regime.
Sympathetic cooling, in which the molecules are allowed to thermalize
with a gas of ultracold atoms, is a promising second-stage cooling
method \cite{Soldan:2004}. However, while elastic collisions allow
thermalization, inelastic collisions can cause trap loss
\cite{deCarvalho:1999}, and for many systems the inelastic collisions
are predicted to be too large for sympathetic cooling to succeed
\cite{Lara:PRA:2007, Zuchowski:NH3:2009, Tokunaga:2011}. Scattering
calculations are essential in order to identify systems for which
sympathetic cooling has a good prospect of success. Once in the
ultracold regime, the extent to which atomic and molecular interactions
can be controlled again depends on a detailed understanding of their
collisional properties.

Quantum molecular scattering calculations are usually carried out using
the coupled-channel method: the Schr\"odinger equation for scattering
is converted into a set of coupled differential equations, which are
then propagated across a range of values of the intermolecular distance
$r$. The size of the problem is determined by the number of channels
$N$ (the number of coupled equations). The usual algorithms take a time
proportional to $N^3$, since each step of the propagation requires an
$\mathcal{O}(N^3)$ matrix operation.

Cold molecule scattering presents problems with a large number of
channels for two reasons:
\begin{enumerate}
\item At very low energies, small splittings between molecular
    energy levels become important. This makes it necessary to
    include fine details of molecular energy level patterns, such
    as tunneling and nuclear hyperfine splitting. The extra degrees
    of freedom require additional basis functions; in particular,
    including nuclear spins can multiply the number of equations by
    a substantial factor (sometimes 100 or more).
\item Collisions in the presence of electric and magnetic fields
    are very important. In an applied field, the total angular
    momentum $J$ is no longer a good quantum number. Because of
    this, the large sets of coupled equations can no longer be
    factorized neatly into smaller blocks for each $J$ as is
    possible in field-free scattering.
\end{enumerate}
In addition, in cold molecule applications it is often necessary to
repeat scattering calculations on a fine grid of energies and/or
applied electric and magnetic fields, which adds greatly to the
computational expense.

Multichannel Quantum Defect Theory (MQDT) offers an alternative to full
coupled-channel calculations. It was originally developed to provide a
uniform treatment of bound and scattering states for problems involving
the interaction an electron with an ion core with Coulomb forces at
long range \cite{Seaton:1966, Seaton:1983}, but was subsequently
generalized to handle a range of other long-range potentials
\cite{Greene:1979, Greene:1982, Seaton:1983, Mies:1984a, Yoo:1986,
Gao:2008}. It has been successfully applied to scattering problems as
diverse as negative ion photodetachment \cite{Watanabe:1980},
near-threshold predissociation of diatomic molecules \cite{Mies:1984a,
Mies:1984} and predissociation of atom-diatom Van der Waals complexes
\cite{Raoult:1988, Raoult:1990}. More recently it has been applied to
ultracold collisions between pairs of neutral atoms
\cite{Julienne:1989, Burke:1998, Mies:MQDT:2000, Raoult:2004,
Gao:2005}, between atoms and ions \cite{Idziaszek:2009, Gao:ion:2010},
and between highly reactive molecules \cite{Idziaszek:PRL:2010,
Idziaszek:PRA:2010, Gao:react:2010}.

MQDT can be viewed in two different ways. The first tries to capture
the important physics of collisions within a few analytic quantum
defect parameters. The other views it as a method for solving the
coupled equations of scattering theory which offers substantial
insights and advantages in efficiency. The common feature of the two
approaches is to take advantage of the enormous difference in energy
and length scales associated with separated collision partners and
short-range potentials.

When MQDT is viewed as a numerical method for solving the coupled
differential equations, the goal is to obtain a matrix $\bm{Y}(E,B)$
\cite{Greene:1982, Mies:1984, Mies:MQDT:2000, Raoult:2004} that
completely describes the short-range dynamics and is insensitive to
collision energy $E$ and magnetic field $B$. This matrix can be
obtained once and then used for calculations over a wide range of
energies and fields, or obtained by interpolation from a few points.
MQDT achieves this by defining $\bm{Y}(E,B)$ at relatively short range,
as described below. The threshold behavior is accounted for from
properties of single channels. Once the matrix $\bm{Y}(E,B)$ has been
obtained, the time required for calculations at additional energies and
fields is only proportional to $N$, not $N^3$.

Understanding threshold atomic physics in quantum defect terms is well
developed \cite{Julienne:1989, Burke:1998, Mies:MQDT:2000, Vogels:2000,
Gao:2001}. Threshold bound-state and scattering properties are
determined mainly by the long-range potential, which can often be
approximated as $-C_n/r^n$. For the case of the Van der Waals
interaction, $-C_6/r^6$, the linearly independent pair of solutions for
a single potential is known \cite{Gao:C6:1998}. An analytic approach to
MQDT using these solutions has been developed \cite{Gao:QDT:1998,
Gao:2009} and gives much insight into ultracold atom-atom collisions
\cite{Julienne:2006}.

This paper investigates the use of MQDT as a numerical method to study
cold atom-molecule collisions. The structure of the paper is as
follows. In Section II we give an overview of the theory of MQDT,
sufficient to define notation. In Section III we apply MQDT to the
prototype system Mg+NH, and compare it with full coupled-channel
calculations in order to establish what is required for it to give
accurate results. In Section IV we present our conclusions and suggest
directions for future work.

\section{Theory}

\subsection{Coupled-channel method}

Cold atomic and molecular collisions and near-threshold bound states
are conveniently described by a set of coupled equations. The
Hamiltonian for an interacting pair of atoms or molecules is of the
form
\begin{equation}
-\frac{\hbar^2}{2\mu} \nabla^2 + \hat{H}_{\text{int}}(\tau) + V(r,\tau),
\end{equation}
where $\mu$ is the reduced mass, $\nabla^2$ is the Laplacian for the
intermolecular coordinates, and $\tau$ denotes all coordinates except
the interparticle distance $r$. $\hat{H}_{\text{int}}(\tau)$ represents
the internal Hamiltonians of the two particles and $V(r,\tau)$ is the
interaction potential. The total wavefunction is expanded
\begin{equation}
\Psi(r,\tau) = r^{-1}\sum_i \varphi_i(\tau)\psi_{i}(r),
\end{equation}
where the $N$ functions $\varphi_i(\tau)$ form a basis set for the
motion in all coordinates, $\tau$, except the intermolecular distance,
and $\psi_{i}(r)$ is the radial wavefunction in channel $i$.
Substituting this expansion into the total time-independent
Schr\"odinger equation and projecting onto the basis function
$\varphi_j(\tau)$ yields the usual coupled equations of scattering
theory,
\begin{equation}\label{eqn:coupledeqn}
\left[-\frac{\hbar^2}{2\mu}\frac{d^2 }{d r^2} -E \right] \psi_{j}(r)
= - \sum_i W_{ji}(r)\psi_{i}(r),
\end{equation}
where $E$ is the energy. The coupling matrix $\bm{W}$ has elements
\begin{eqnarray}\label{eqn:W1}
W_{ji}(r) = \int \varphi_j^*(\tau) &&\bigg[H_{\text{int}}(\tau) +
V(r,\tau) \nonumber\\
&&\quad+\ \frac{\hbar^2 L_i(L_i+1)}{2\mu r^2} \bigg]
\varphi_i(\tau)\, d\tau ,
\end{eqnarray}
where $L_i$ is the partial-wave quantum number for channel $i$.
Equation (\ref{eqn:coupledeqn}) can conveniently be written in matrix
form,
\begin{equation}\label{eqn:matrix}
 \frac{\hbar^2}{2\mu} \frac{d^2\bm{\psi}}{dr^2}
 = [\bm{W}(r) - E\bm{I}]\bm{\psi}(r),
\end{equation}
where $\bm{\psi}(r)$ is a column vector made up of the solutions
$\psi_i(r)$ and $\bm{I}$ is the identity matrix.

For both bound states and collision calculations, the wavefunction must
be regular at the origin. When $V(r) \gg 0$ as $r \rightarrow 0$, the
short-range boundary condition is
\begin{equation}\label{eqn:srbc}
\psi_i(r) \rightarrow 0 \text{\quad as\quad} r \rightarrow 0.
\end{equation}
At any energy, there are $N$ linearly independent solution vectors
$\bm{\psi}(r)$ that satisfy these boundary conditions, and it is
convenient to combine them to form the $N \times N$ wavefunction matrix
$\bm{\Psi}$(r).

The coupled-channel approach propagates either the wavefunction matrix
$\bm{\Psi}(r)$ and its derivative $\bm{\Psi}'(r) $, or the
log-derivative matrix $ \bm{L}(r) = \bm{\Psi}' [\bm{\Psi}]^{-1}$,
outwards from $r=0$ (or a point in the deeply classically forbidden
region at short range) \cite{Gordon:1969, Johnson:1973}. In scattering
calculations, the propagation is continued to a point $r_\text{max}$ at
large $r$. The wavefunction or log-derivative matrix is then
transformed into a representation where $\bm{W}$ is asymptotically
diagonal \cite{Gonzalez-Martinez:2007}, such that
\begin{equation}\label{eqn:W}
 W_{ji}(r) \stackrel{r \rightarrow \infty}{\longrightarrow}
 \left[ E_i^\infty + \frac{\hbar^2 L_i(L_i+1)}{2 \mu r^2} \right]\delta_{ij}
 + \mathcal{O}(r^{-n}),
\end{equation}
where $n$ is the power of the leading term in the potential expansion
and $E_i^\infty$ is the threshold of channel $i$. Each channel is
either asymptotically open, $E \ge E_i^\infty$, or asymptotically
closed, $E < E_i^\infty$. The scattering boundary conditions are
\begin{equation}\label{eqn:match}
\bm{\Psi} = \bm{J}(r) + \bm{N}(r)\bm{K}.
\end{equation}
The matrices $\bm{J}$ and $\bm{N}$ are diagonal matrices containing
Riccati-Bessel functions for open channels and modified spherical
Bessel functions for closed channels \cite{Johnson:1973}. In a problem
containing $N$ channels, $N_{\text{o}}$ of which are open, the
scattering $\bm{S}$ matrix is related to the open-open submatrix of
$\bm{K}$ by
\begin{equation}\label{eqn:S}
 \bm{S} = (1 + i\bm{K}_{\text{oo}})^{-1}(1-i\bm{K}_{\text{oo}}).
\end{equation}
In full coupled-channel calculations, the matrices $\bm{K}$ and
$\bm{S}$ are rapidly changing functions of both energy and field,
particularly near scattering resonances, so that the entire propagation
to long range must be repeated for each set of conditions required.

\subsection{Multichannel Quantum Defect Theory}

MQDT also begins by propagating the wavefunction or log-derivative
matrix outwards from short range. However, instead of continuing to
$r_\text{max}$, matching takes place at a point $r_\text{match}$, at
relatively short range. The matching in MQDT treats the open and weakly
closed channels on an equal footing; weakly closed channels are usually
defined as those that are locally open, $E>W_{ii}(r)$, at some value of
$r$, so are capable of supporting scattering resonances. Matching at
short range produces a matrix $\bm{Y}(E,B)$ that is relatively
insensitive to energy and applied field, as described below. $\bm{Y}$
also varies smoothly across thresholds, unlike $\bm{S}$ and $\bm{K}$.
Provided the channels are uncoupled outside $r_\text{match}$, it is
then possible to obtain the scattering $\bm{S}$ matrix from $\bm{Y}$
using the properties of individual uncoupled channels.

We consider a problem with $N_\text{o}$ open channels and $N_\text{c}$
weakly closed channels at some collision energy $E$ and field $B$. For
each such channel, $i =1,N$, where $N=N_\text{o}+N_\text{c}$, MQDT
requires a reference potential, $U_i^{\text{ref}}(r)$, which
asymptotically has similar behavior to $W_{ii}(r)$ in equation
(\ref{eqn:W}). This reference potential defines a linearly independent
pair of reference functions $f_i(r)$ and $g_i(r)$,
\begin{align}\label{eqn:Uref}
\left[ \frac{d^2}{dr^2} + K^2_i(r)\right]f_i(r) = 0,
\end{align}
and similarly for $g_i$, where the local wave vector $K_i(r)$ is
\begin{align}
K_i(r)=\sqrt{\frac{2\mu}{\hbar^2}(E-U^{\text{ref}}_i(r))}.
\end{align}
The regular solution $f_i$ has the boundary condition $f_i \rightarrow
0$ as $r \rightarrow 0$. $f_i$ and $g_i$ are normalized to have
Wentzel-Kramers-Brillouin (WKB) form, with amplitude $K_i(r)^{-1/2}$,
at some point in the classically allowed region \cite{Greene:1982}. The
$N\times N$ matrix $\bm{Y}$ is defined by matching at $r_\text{match}$,
\begin{equation}\label{eqn:match_mqdt}
 \bm{\Psi} = \bm{f}(r) + \bm{g}(r)\bm{Y},
\end{equation}
or in terms of the log-derivative matrix $\bm{L}$,
\begin{equation}
(\bm{Lf}-\bm{f}') = (\bm{Lg}-\bm{g'})\bm{Y},
\end{equation}
where $\bm{f}$ and $\bm{g}$ are diagonal matrices containing the
functions $f_i$ and $g_i$ and the primes indicate radial derivatives.

In order to relate $\bm{Y}$ to the physical scattering $\bm{S}$ matrix,
the asymptotic forms of the reference functions $f_i$ and $g_i$ in each
channel are required. To this end another pair of reference functions
is defined for each channel. For open channels, these functions are
asymptotically energy-normalized,
\begin{align}\label{eqn:sc}
s_i(r) \stackrel{r \rightarrow \infty}{\longrightarrow} k_i^{-\frac{1}{2}}
\sin{ \left( k_ir - \frac{L_i\pi}{2} + \xi_i\right)}, \\ \label{eqn:s}
c_i(r) \stackrel{r \rightarrow \infty}{\longrightarrow} k_i^{-\frac{1}{2}}
\cos{ \left( k_ir - \frac{L_i\pi}{2} + \xi_i\right)},
\end{align}
where $\xi_i$ is the phase shift associated with reference potential
$i$ and $k_i$ is the asymptotic wave vector,
\begin{equation}
 k_i = \sqrt{\frac{2 \mu}{\hbar^2} (E-E_i^\infty)}.
 \end{equation}
These asymptotically normalized functions are related to $f_i$ and
$g_i$ through the quantum defect parameters $C_i$ and $\tan\lambda_i$,
\begin{eqnarray}
s_i(r) &=& C_i^{-1}f_i(r);\label{eqn:fC}  \\
c_i(r) &=& C_i[g_i(r) + \tan\lambda_i f_i(r)].
\label{eqn:gC}
\end{eqnarray}
Thus $C_i$ relates the amplitudes of the energy-normalised reference
functions to WKB-normalised ones, while $\tan\lambda_i$ describes the
modification in phase due to threshold effects. Far from threshold,
$C_i\approx 1$ and $\tan\lambda_i\approx 0$.

For each weakly closed channel, an exponentially decaying solution is
defined,
\begin{equation}\label{eqn:phi}
\phi_i(r) \stackrel{r \rightarrow \infty}{\longrightarrow}
\textstyle{\frac{1}{2}}e^{-|k_i|r} \sqrt{|k_i|}.
\end{equation}
This is related to the solutions $f_i$ and $g_i$ by a normalization
factor $\mathcal{N}_i$, and an energy-dependent phase $\nu_i$,
\begin{align} \phi_i(r)
= \mathcal{N}_i \left[\cos{ \nu}_i f_i(r) -\sin{
\nu}_i g_i(r)\right].\label{eqn:phi1}
\end{align}
The phase $\nu_i$ is an integer multiple of $\pi$ at each energy that
corresponds to a bound state of the reference potential in channel $i$.

The $\bm{Y}$ matrix is converted into the $\bm{S}$ matrix of scattering
theory using the quantum defect parameters $C_i$, $\tan\lambda_i$,
$\tan\nu_i$ and $\xi_i$. First, the effect of coupling to closed
channels is accounted for,
\begin{equation}\label{eqn:ybar}
\bm{\overline{Y}} = \bm{Y}_{\text{oo}}-\bm{Y}_{\text{oc}}
[\tan\bm{\nu} + \bm{Y}_{\text{cc}}]^{-1}\bm{Y}_{\text{co}},
\end{equation}
where $\tan\bm{\nu}$ is a diagonal matrix of dimension $N_{\text{c}}
\times N_{\text{c}}$ containing elements $\tan\nu_i$. The
$N_\text{o}\times N_\text{o}$ matrix $\bm{\overline{Y}}$ incorporates
any resonance structure caused by coupling to closed channels through
$\tan\bm{\nu}$. Unlike $\bm{Y}$ itself, $\bm{\overline{Y}}$ can be a
rapidly varying function of energy and field. Secondly, threshold
effects from asymptotically open channels are incorporated,
\begin{equation}\label{eqn:RY}
\bm{\overline{R}} = \bm{C}^{-1}
\left[\bm{\overline{Y}}^{-1}-\tan\bm{\lambda}\right]^{-1}\bm{C}^{-1},
\end{equation}
where $\bm{C}$ and $\tan\bm{\lambda}$ are diagonal matrices of
dimension $N_{\text{o}} \times N_\text{o}$, containing elements $C_i$
and $\tan\lambda_i$. Finally, the $\bm{S}$ matrix is obtained from
\begin{equation}\label{eqn:SR}
\bm{S} = e^{i\bm{\xi}} \left[1 + i\bm{\overline{R}}\right]
\left[1 - i\bm{\overline{R}}\right]^{-1}e^{i\bm{\xi}}.
\end{equation}
This may be compared to equation (\ref{eqn:S}) for the full
coupled-channel method. The inclusion of the diagonal matrix
$e^{i\bm{\xi}}$ accounts for the phase difference between the reference
functions $f_i$ and $g_i$ used by MQDT and the Riccati-Bessel functions
used by the full coupled-channel method.

The approach taken in the present paper is somewhat different from that
in refs.\ \cite{Mies:1984a, Mies:MQDT:2000}. There MQDT was approached
as an exact representation of the full coupled-channel solution. The
matrix $\bm{Y}$ was evaluated at a distance $r_\text{match}$ large
enough that it had become constant as a function of $r_\text{match}$.
When this is done, MQDT gives the same (exact) results for any choice
of reference potential $U^{\rm ref}_i(r)$, although constancy of
$\bm{Y}$ may be achieved at different values of $r_\text{match}$ for
different choices. In our approach, $r_\text{match}$ is chosen to
ensure that $\bm{Y}$ is only weakly energy-dependent, and this may
require matching in a region where $\bm{Y}$ is not yet independent of
$r_\text{match}$. With this approach, MQDT provides an {\em
approximate} solution whose quality depends on the choice of reference
potentials.

\subsection{Numerical evaluation of reference functions
and quantum defect parameters}\label{sec:reffn}

\subsubsection{Open channels}

For an open channel $i$, the reference function $s_i$ is obtained by
propagating a regular solution of (\ref{eqn:Uref}) from a point inside
$r_\text{match}$ to a point $r_\text{max}$ at long range and imposing
the boundary condition (\ref{eqn:sc}) (or its Bessel function
equivalent). This establishes the normalization of $s_i$ and also gives
the phase shift $\xi_i$, which is then used to obtain the function
$c_i$ at $r_\text{max}$ from the boundary condition (\ref{eqn:s}). The
reference function $c_i$ is then propagated inwards to
$r_\text{match}$. The two remaining quantum defect parameters are
obtained by applying \cite{Mies:1984a}
\begin{equation}
C_i^{-2} = (s_i^2 K_i + s_i'^2/K_i)
\end{equation}
and
\begin{equation}
\cot\lambda_i=\frac{K_i(\gamma_i-u_i)}{K_i^2+\gamma_i u_i}
\end{equation}
in the classically allowed region, where $\gamma_i = s_i'/s_i$ and
$u_i=c_i'/c_i$. The primes indicate radial derivatives. Equations
(\ref{eqn:fC}) and (\ref{eqn:gC}) then give the reference functions
$f_i$ and $g_i$.

\subsubsection{Closed channels}

For a weakly closed channel $i$, the reference function $f_i$ is again
obtained by propagating a regular solution of (\ref{eqn:Uref}) outwards
from a point inside $r_\text{match}$, but in this case $f_i$ is
normalized in the classically allowed region such that
\begin{equation}\label{eqn:norm}
f_i^2(K_i^2+\gamma_i^2) = K_i.
\end{equation}
In the closed-channel case, $g_i$ cannot be obtained directly from
$f_i$ at a single point. Instead, the reference function $\phi_i$ is
obtained by using (\ref{eqn:phi}) as a long-range boundary condition
and propagating a solution of (\ref{eqn:Uref}) inwards towards $r=0$.
The normalization factor $\mathcal{N}_i$ of equation (\ref{eqn:phi1})
is obtained by matching to
\begin{equation}
\mathcal{N}_i^2=(\phi_i^2K_i+\phi_i'^2/K_i).
\end{equation}
in the classically allowed region. The quantum defect parameter
$\tan\nu_i$ is then obtained from
\begin{equation}\label{eqn:nu}
\tan\nu_i= \frac{K_i(t_i-\gamma_i)}{K_i^2+\gamma_i t_i},
\end{equation}
where $t_i = \phi_i'/\phi_i$. Finally, the function $g_i$ is obtained
from $f_i$ and $\phi_i$ using equation (\ref{eqn:phi1}).

%Further details on the normalizations used in this section can be found
%in \cite{0034-4885-46-2-002}.

\subsection{Sources of Error}

There are a number of sources of errors in MQDT calculations using our
approach:
\begin{enumerate}
\item Interchannel couplings that occur outside $r_\text{match}$,
    which are not taken into account by equations (\ref{eqn:ybar})
    to (\ref{eqn:SR});
\item Deviations between the reference potentials
    $U_i^{\text{ref}}(r)$ and $W_{ii}(r)$ outside $r_\text{match}$;
\item Differences between the actual $\bm{Y}$ matrix at a given
    energy and field and the $\bm{Y}$ matrix obtained by
    interpolation.
\end{enumerate}

\section{Results and Discussion}

To explore the application of MQDT to cold molecular collisions, we
consider the prototype system Mg+NH($^3\Sigma^-$). The potential energy
surface for this system is moderately anisotropic
\cite{Soldan:MgNH:2009} and provides substantial coupling between
channels. The system is topical because Wallis and Hutson
\cite{Wallis:MgNH:2009} have shown that sympathetic cooling of cold NH
molecules by ultracold Mg atoms has a good prospect of success.

The energy levels of NH in a magnetic field are most conveniently
described using Hund's case (b), in which the molecular rotation $n$
couples to the spin $s$ to produce a total monomer angular momentum
$j$. In zero field, each rotational level $n$ is split into sublevels
labeled by $j$. In a magnetic field, each sublevel splits further into
$2j + 1$ levels labeled by $m_j$, the projection of $j$ onto the axis
defined by the field. For the $n = 0$ levels that are of most interest
for cold molecule studies, there is only a single zero-field level with
$j = 1$ that splits into three components with $m_j = +1$, $0$ and
$-1$.

The coupled equations are constructed in a partly coupled basis set
$|nsjm_j \rangle |LM_L \rangle$, where $L$ is the end-over-end
rotational angular momentum of the Mg atom and the NH molecule about
one another and $M_L$ is its projection on the axis defined by the
magnetic field. Hyperfine structure is neglected. The matrix elements
of the total Hamiltonian in this basis are given in ref.\
\cite{Gonzalez-Martinez:2007}. The only good quantum numbers during the
collision are the parity $p = (-1)^{n+L+1}$ and the total projection
quantum number $M = m_j + M_L$. The calculations in the present work
are performed for $p=-1$ and $M = 1$. This choice includes s-wave
scattering of NH molecules in initial state $m_j = +1$, which is
magnetically trappable, to $m_j = 0$ and $-1$, which are not. The basis
set used included all functions up to $n_{\text{max}} = 1$ and
$L_{\text{max}} = 3$. This unconverged basis set is sufficient for the
purpose of comparing MQDT results with full coupled-channel
calculations.

\subsection{Numerical methods}

The coupled-channel calculations required for both MQDT and the full
coupled-channel approach were carried out using the MOLSCAT package
\cite{molscat:v15}, as modified to handle collisions in magnetic fields
\cite{Gonzalez-Martinez:2007}. The coupled equations were solved
numerically using the hybrid log-derivative propagator of Alexander and
Manolopoulos \cite{Alexander:1987}, which uses a fixed-step-size
log-derivative propagator in the short-range region ($r_\text{min} \le
r < r_\text{mid}$) and a variable-step-size Airy propagator in
long-range region ($r_\text{mid} \le r \le r_\text{max}$. The full
coupled-channel calculations used $r_\text{min}=2.5$~\AA,
$r_\text{mid}=50$~\AA\ and $r_\text{max}=250$~\AA\ (where 1~\AA\ =
$10^{-10}$~m). MQDT requires coupled-channel calculations only from
$r_\text{min}$ to $r_{\rm match}$ (which is less than $r_\text{mid}$),
so only the fixed-step-size propagator was used in this case.

The MQDT reference functions and quantum defect parameters were
obtained as described in Section \ref{sec:reffn}, using the Numerov
propagator \cite{Numerov:1933} to solve the 1-dimension Schr\"odinger
equations. Use of the renormalized Numerov method \cite{Johnson:1977}
was not found to be necessary in the present case. The MQDT $\bm{Y}$
matrix was then obtained by matching to the log-derivative matrix
extracted from the coupled-channel propagation at a distance
$r_\text{match}$.

\begin{figure}[tb]
\centering
\includegraphics[width=1.0\columnwidth,clip]{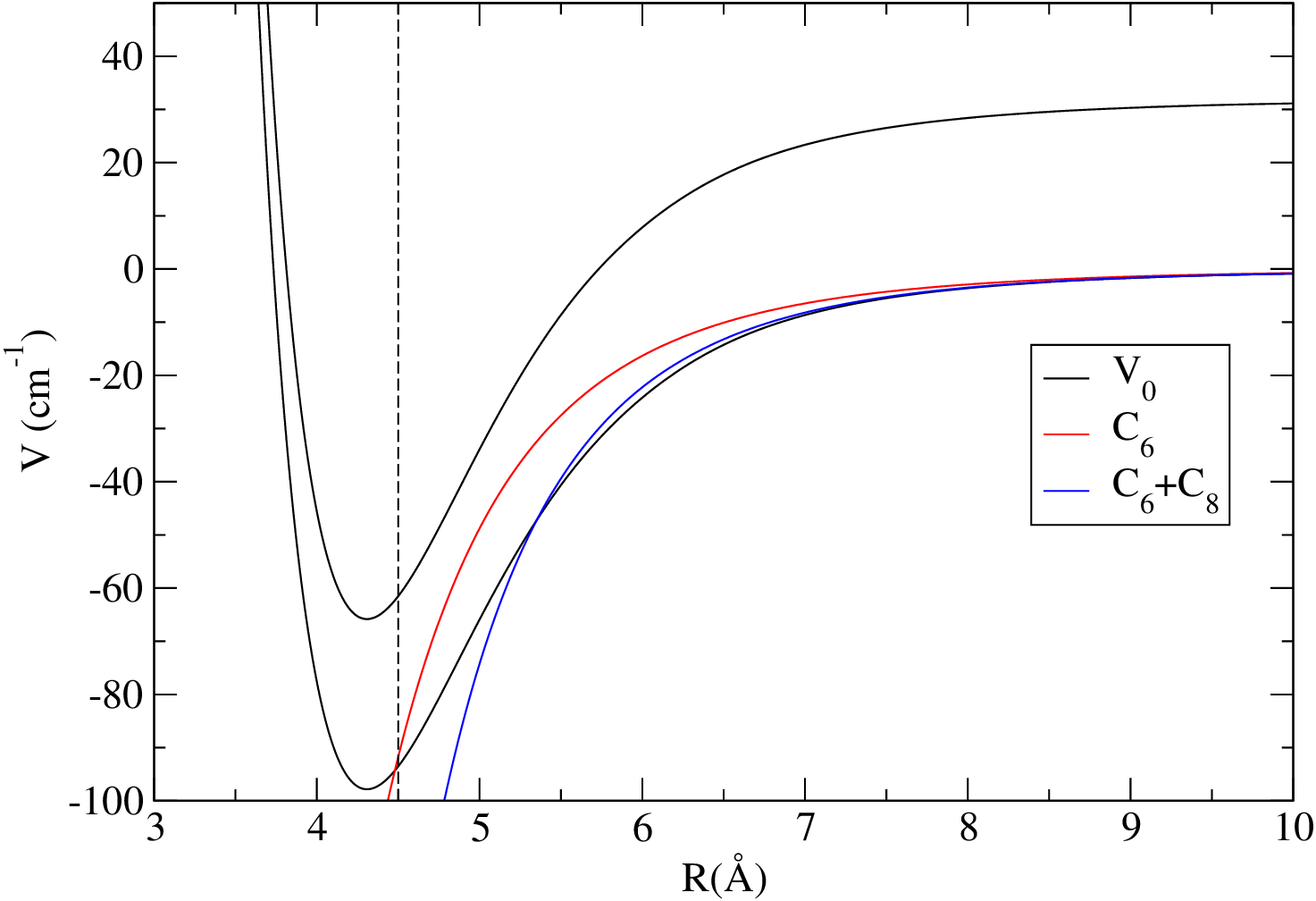}
\caption{(Color online) Zero-field reference potentials. For the $V_0$
reference potential the first rotational excited state is also shown
($n=1$). The hard wall at $r=4.5$~\AA\ is shown as a dashed line.}
\label{figure:pot}
 % T.eps: 0x0 pixel, 300dpi, 0.00x0.00 cm, bb=(atend)
\end{figure}

%%%%%%%%%%%%%%%%%%%%%%%%%%%%%%%%%%%%%%%%%%%%%%%%%%%%%%%%%%%%%%%%%%%%%%
\subsection{Comparison of full coupled-channel and MQDT results}

\subsubsection{Choice of $r_{\rm match}$ and reference potential}
\label{sec:refchoice}

One of the goals of MQDT is to obtain a matrix $\bm{Y}(E,B)$ in such a
way that it is only weakly dependent on energy $E$ and magnetic field
$B$. However, the actual form of $\bm{Y}(E,B)$ is strongly dependent on
the distance at which it is defined and the reference potentials used.
In the present work we consider three different reference potentials,
as shown in Figure \ref{figure:pot}. First we define a reference
potential containing a pure $C_6$ long-range term, which has been used
with great success in cold atom-atom collisions,
\begin{equation}\label{eqn:refunc1}
U^{\text{ref},C_6}_i(r) =
-\frac{C_6}{r^6} + \frac{\hbar^2 L_i(L_i+1)}{2\mu r^2}
+ E_i^\infty,
\end{equation}
where $C_6=7.621 \times 10^5 \text{ \AA}^6\text{ cm}^{-1}$ for Mg+NH
\cite{Soldan:MgNH:2009}. Secondly we define a reference potential
containing an additional $C_8$ term,
\begin{equation}\label{eqn:refunc2}
U^{\text{ref},C_{6,8}}_i(r) =
-\frac{C_6}{r^6} - \frac{C_8}{r^8} + \frac{\hbar^2 L_i(L_i+1)}{2\mu r^2}
+ E_i^\infty,
\end{equation}
where $C_8=9.941 \times 10^6 \text{ \AA}^8\text{ cm}^{-1}$
\cite{Soldan:MgNH:2009}. Finally we define
\begin{equation}\label{eqn:refunc3}
U^{\text{ref},V_0}_i(r) =
V_0(r) + \frac{\hbar^2 L_i(L_i+1)}{2\mu r^2} + E_i^\infty,
\end{equation}
where $V_0(r)$ is the isotropic part of the interaction potential,
which is equivalent to the diagonal $\bm{W}$ matrix element in the
incoming s-wave channel. Each reference potential contains a hard wall
at $r = r^{\text{wall}}_i$, so that $U^\text{ref}_i(r) = \infty$ for
$r<r^{\text{wall}}_i$. This allows the phase $\xi_i$ of the reference
functions in each channel to be adjusted if required. A useful feature
of MQDT, to be explored in future work, is that the position of the
hard wall can be chosen to minimize the energy-dependence of $\bm{Y}$.
However, in the present paper we simply take $r^{\text{wall}}_i =
4.5$~\AA.

It is convenient to compare MQDT and coupled-channel results at the
level of T-matrix elements, $T_{ij}=\delta_{ij}-S_{ij}$. In general we
label elements $T_{ \alpha,L,M_L \rightarrow \alpha',L',M_L'}$, where
$|\alpha \rangle = |nsjm_j \rangle$. However, the collisions considered
in the present paper are all among the $n=0, j=1$ levels and so
$\alpha$ is simply abbreviated to $m_j$. The spin-changing cross
sections are quite small except near resonances, so we focus mostly on
diagonal elements, for which we suppress the second set of labels.

\begin{figure}[tb]
\centering
\includegraphics[width=1.0\columnwidth,clip]{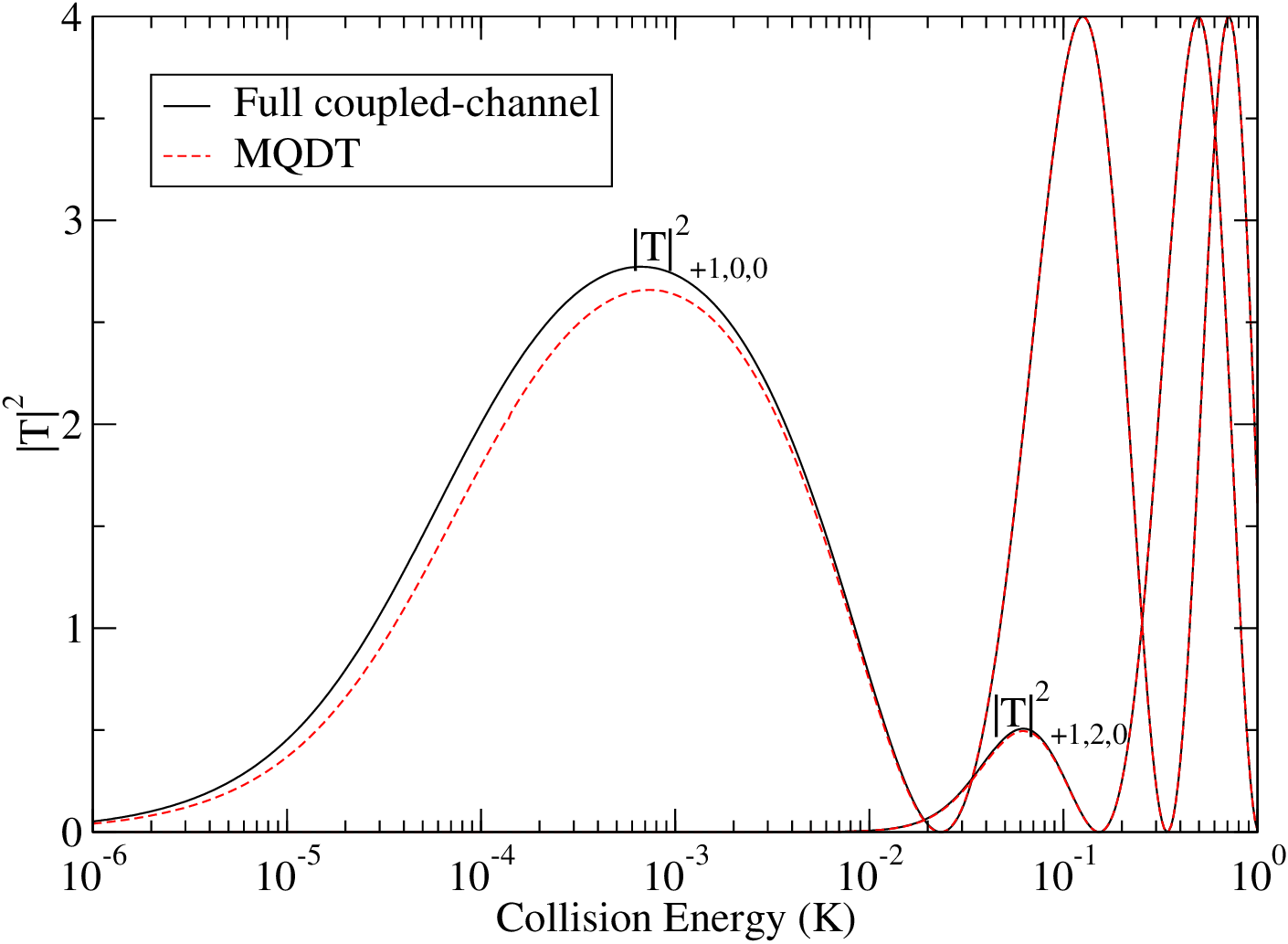}
\caption{(Color online)
The squares of diagonal $T$-matrix elements in the incoming channels
for $m_j=+1$, $L=0$ and 2 at $B=10$~G, obtained from full
coupled-channel calculations (solid, black) and MQDT with the $C_6$
reference potential and $r_\text{match}=20$~\AA\ (dashed, red).
$T$-matrix elements are labeled with quantum numbers $m_j,L,M_L$.
[Units of Gauss rather than Tesla, the accepted SI unit of magnetic
field, have been used in this paper to conform to the conventional
usage of this field.]} \label{figure:T_C6_20}
% T.eps: 0x0 pixel, 300dpi, 0.00x0.00 cm, bb=(atend)
\end{figure}

Figure \ref{figure:T_C6_20} compares diagonal T-matrix elements
$|T_{ii}|^2$ obtained from full coupled-channel calculations with those
from the MQDT method for the pure $C_6$ reference potential of equation
(\ref{eqn:refunc1}), with a matching distance of $r_\text{match}=20$
\AA. The $\bm{Y}$ matrix was recalculated at every energy at which full
coupled-channel calculations were performed. The MQDT results reproduce
the coupled-channel results almost exactly at collision energies
$E/k_{\rm B} > 10$~mK. However, at lower energies the results start to
differ noticeably. It may be noted that $|U_i^{\text{ref},C_6} -
W_{ii}|/k_{\rm B}\approx 0.6$~mK at $r_\text{match}=20$~\AA.
%The overall conclusion is that MQDT is reliable only when $ E\gg
%|U_i^{\text{ref}} - W_{ii}|$ in all channels.

\begin{figure}[tb]
\centering
\includegraphics[width=1.0\columnwidth,clip]{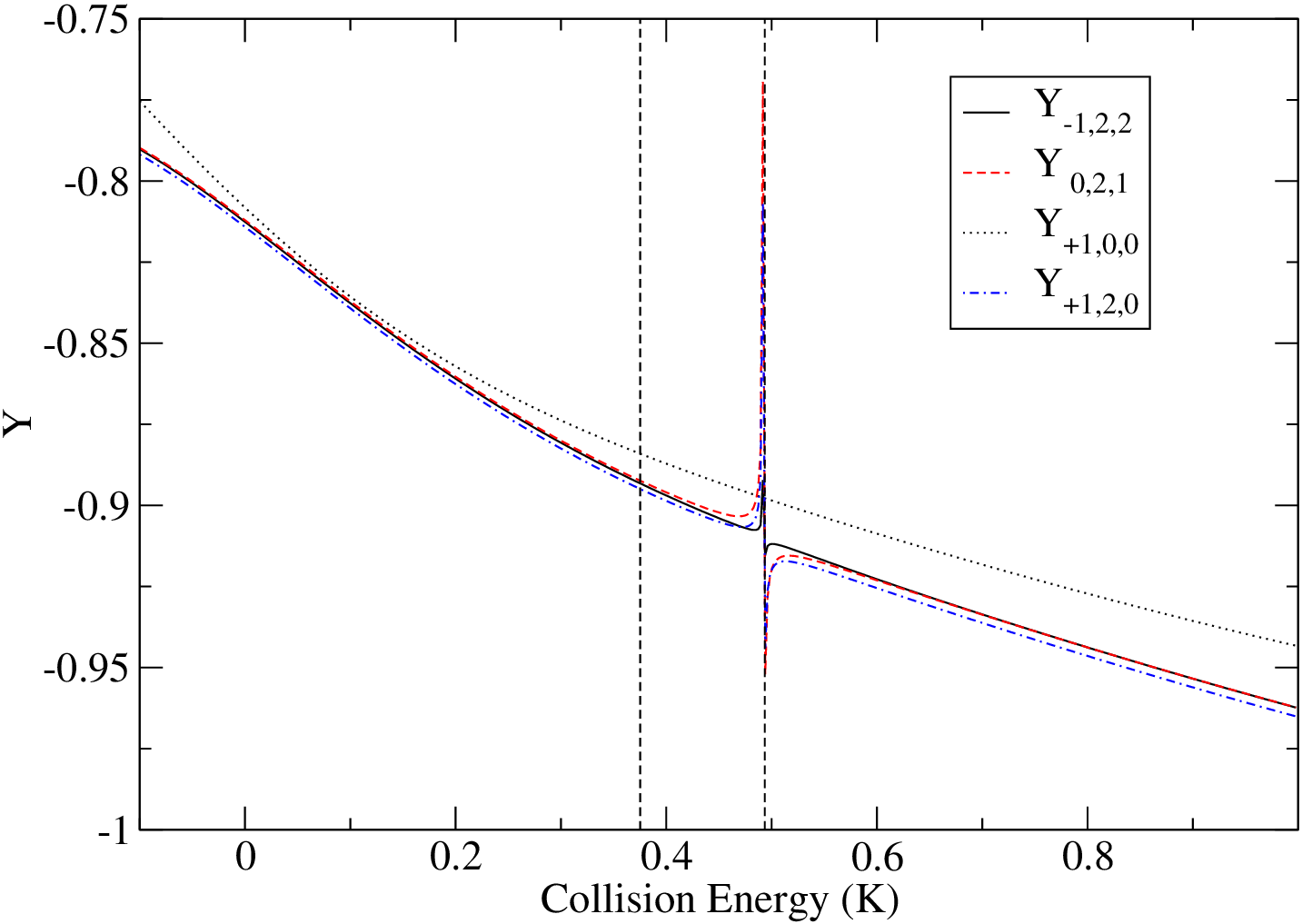}
\caption{(Color online) Diagonal $\bm{Y}$ matrix elements as a function
of collision energy  at $B=10$~G for the $C_6$ reference potential with
$r_\text{match} = 20$~\AA. The dashed vertical lines show the positions
of quasibound states as described in the text.}
 \label{figure:Y_C6_20}
 % T.eps: 0x0 pixel, 300dpi, 0.00x0.00 cm, bb=(atend)
\end{figure}

Figure \ref{figure:Y_C6_20} shows the diagonal $\bm{Y}$ elements
corresponding to Figure \ref{figure:T_C6_20}. They vary smoothly across
most of the energy range, and are continuous across the threshold at
zero energy, but exhibit occasional sharp structures as a function of
energy. These sharp features are close to the energies of quasibound
states, as shown by carrying out bound-state calculations using the
BOUND package \cite{Hutson:bound:1993}, with the same basis set as the
MOLSCAT calculations. The resulting bound-state energies are shown in
Figure \ref{figure:Y_C6_20} as dashed vertical lines. The broad feature
near $E/k_{\rm B}=0.5$~K is due to a quasibound state (Feshbach
resonance) with quantum numbers $n=1$, $j=0$, $m_j=0$, $L=3$.

For MQDT to be more efficient than full coupled-channel calculations,
it needs to produce results in agreement with full coupled-channel
calculations from an energy-insensitive $\bm{Y}$ matrix that can be
assumed to be constant or can be obtained by interpolation from a few
energies, instead of being recalculated at every energy. However, the
$\bm{Y}$ matrix elements in Figure \ref{figure:Y_C6_20} do not meet
this requirement: the resonant features prevent reliable interpolation
over useful ranges of energy.

\begin{figure}[tb]
\centering
\includegraphics[width=1.0\columnwidth,clip]{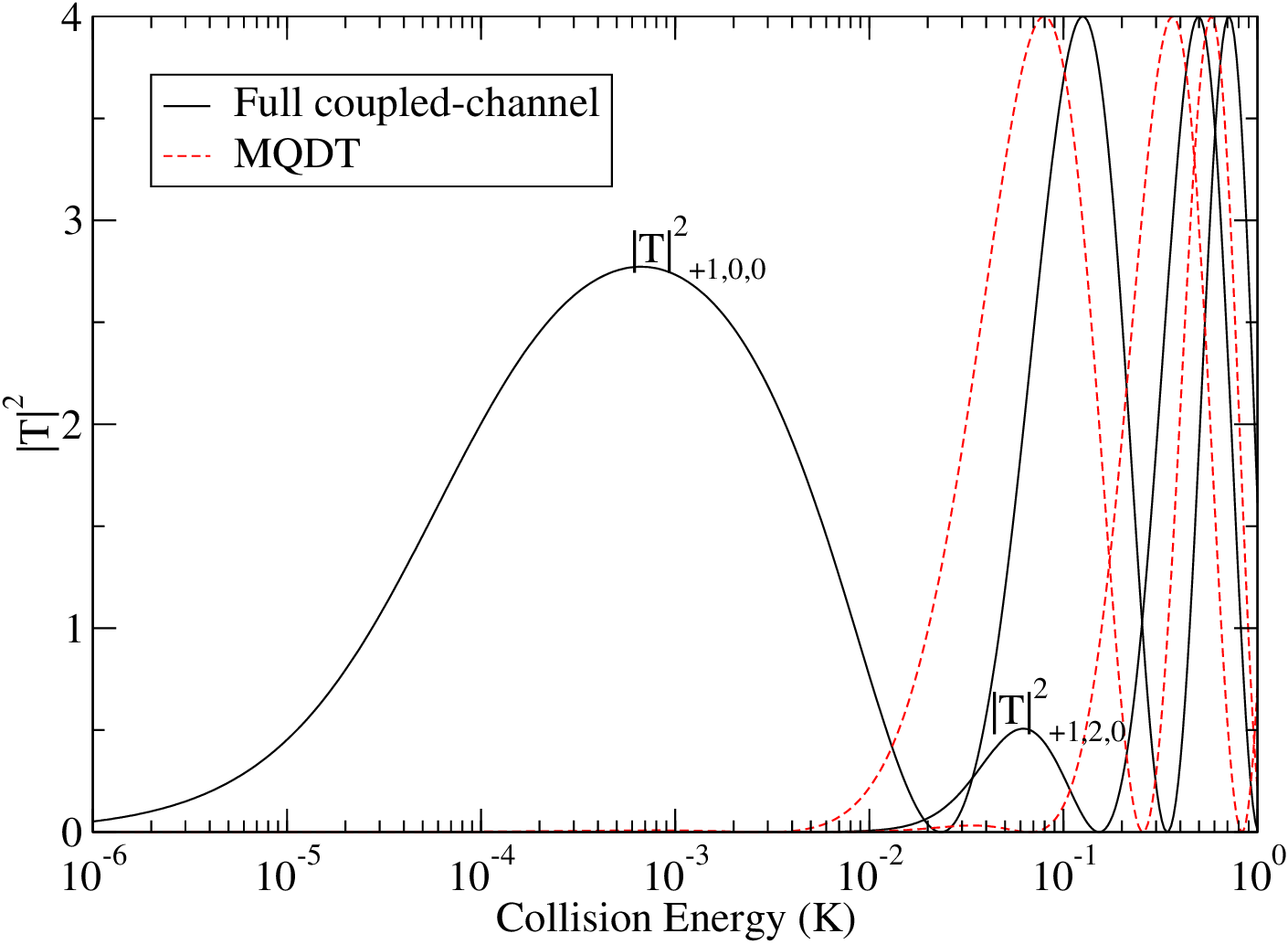}
\caption{(Color online)
The squares of diagonal $T$-matrix elements in the incoming channels
for $m_j=+1$, $L=0$ and 2 at $B=10$~G, obtained from full
coupled-channel calculations (solid, black) and MQDT with the $C_6$
reference potential and $r_\text{match}=6.8$~\AA\ (dashed, red).}
\label{figure:T_68_C6}
% T.eps: 0x0 pixel, 300dpi, 0.00x0.00 cm, bb=(atend)
\end{figure}

The energy sensitivity of the $\bm{Y}$ matrix in Figure
\ref{figure:Y_C6_20} is due to the value used for $r_\text{match}$.
When $r_\text{match}$ is large, resonance features due to quasibound
states may be present in the log-derivative matrix from which $\bm{Y}$
is obtained. In this case the open and closed-channel blocks of
$\bm{Y}$ are uncoupled, so that $\bar{\bm{Y}} \approx \bm{Y}_{\rm oo}$,
and the resonances appear through the $\bm{Y}_{\rm oo}$ term in Eq.\
\ref{eqn:ybar} rather than through $\tan\bm{\nu}+\bm{Y}_{\rm cc}$
\cite{Fourre:1994}. However, if $r_\text{match}$ is small enough, the
resonance features are shifted to high energies, out of the region of
interest. It is usually desirable to obtain $\bm{Y}$ at a value of
$r_\text{match}$ that is in or near the classically allowed region for
all weakly closed channels. However it must be remembered that the MQDT
method neglects interchannel couplings that occur outside
$r_\text{match}$, so there is always a tradeoff between choosing a
value that minimizes the energy-dependence and one that takes account
of coupling at relatively long range. This is particularly important in
molecular scattering, where the anisotropy of the interaction potential
often provides substantial couplings at long range.

It is convenient to consider lengths and energies in ultracold
scattering in relation to the Van der Waals characteristic length and
energy, defined by \cite{Jones:RMP:2006}
\begin{equation}
r_\text{VdW}=\frac{1}{2}\left(\frac{2\mu C_6}{\hbar^2}\right)^\frac{1}{4}
\text{\quad and \quad} E_{\rm VdW} = \frac{\hbar^2}{2\mu r_\text{VdW}^2}.
\end{equation}
For Mg+NH, $r_\text{VdW}=12.7$~\AA\ and $E_\text{VdW}/k_{\rm B}=11$~mK.
In atomic systems, it is common to place $r_\text{match}$ close to
$r_\text{VdW}$. However, the quasibound state responsible for the broad
feature in Figure \ref{figure:Y_C6_20} is due to an $n=1$ state, with
an outer turning point around 5.7~\AA. The resonant feature therefore
does not shift in energy significantly until $r_\text{match}$ is around
7~\AA. In addition, it is not enough simply to move $r_\text{match}$ to
short range with the same reference function. Figure
\ref{figure:T_68_C6} shows diagonal T-matrix elements obtained by MQDT
with the $C_6$ reference function, as in Figure \ref{figure:T_C6_20},
but with $r_\text{match}=6.8$~\AA. This does indeed produce a $\bm{Y}$
matrix without poles in the energy region of interest, but the MQDT
results are no longer in agreement with the full coupled-channel
results at any of the energies considered. This is because the
difference between the reference potential and the diagonal $\bm{W}$
matrix elements at $r_\text{match}=6.8$~\AA\ is $| U_i^{\text{ref},C_6}
- W_{ii}|/k_{\rm B}\approx 4$~K, as seen in Figure \ref{figure:pot}.
Alternatively, in terms of the approach of Mies and Raoult
\cite{Mies:MQDT:2000}, 6.8~\AA\ is too short a distance for the
$\bm{Y}$ matrix evaluated with the $C_6$ reference potential to have
reached its asymptotic value.

\begin{figure}[tb]
\centering
\includegraphics[width=1.0\columnwidth,clip]{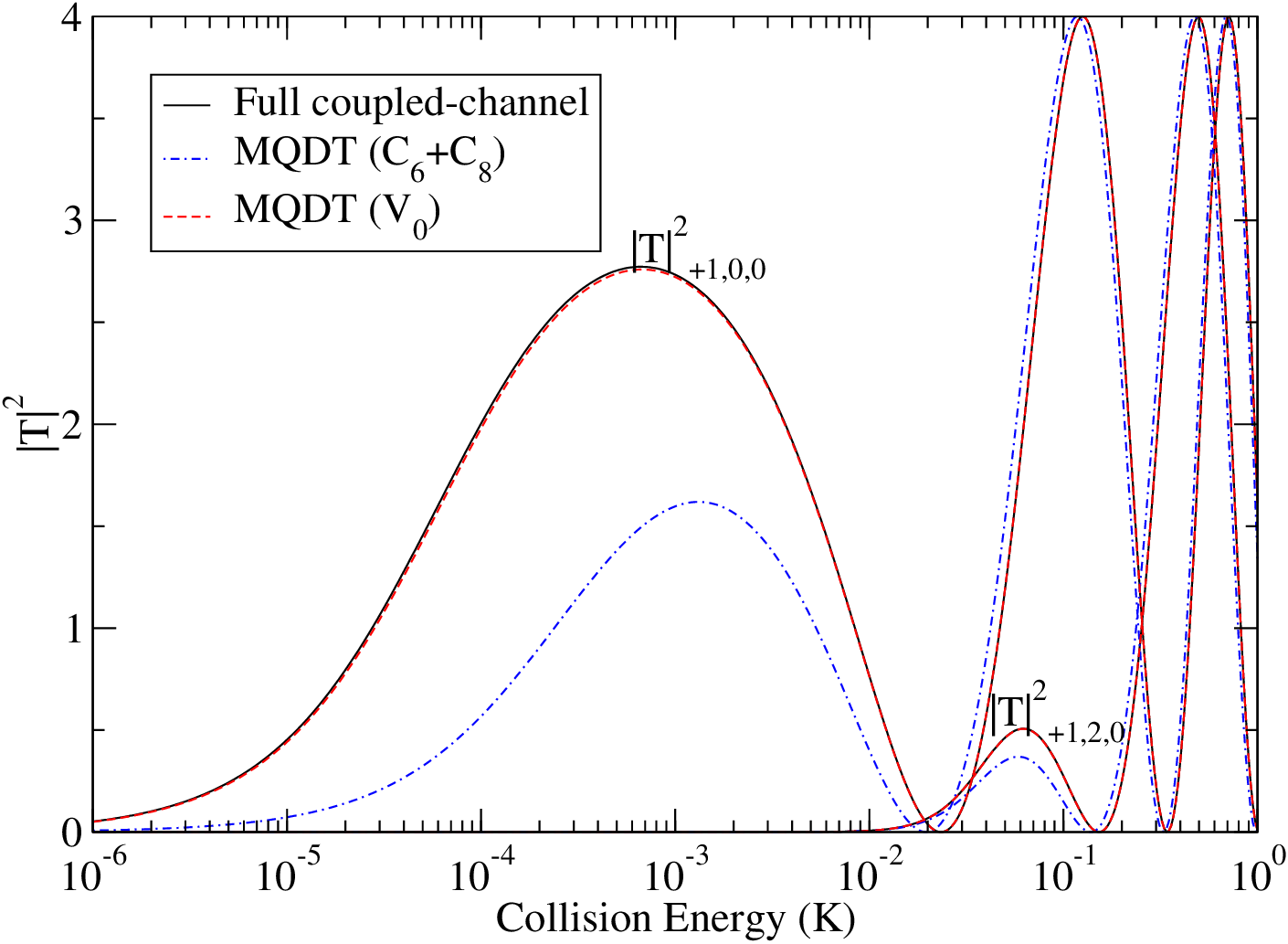}
\caption{(Color online)
The squares of diagonal $T$-matrix elements in the incoming channels
for $m_j=+1$, $L=0$ and 2 at $B=10$~G, obtained from full
coupled-channel calculations (solid, black) and MQDT with the $C_6+C_8$
(dot-dashed, blue) and $V_0$ (dashed, red) reference potentials and
$r_\text{match} =6.8$ \AA.} \label{figure:T_68_68V0}
% T.eps: 0x0 pixel, 300dpi, 0.00x0.00 cm, bb=(atend)
\end{figure}

This problem may be remedied by using a better reference potential.
Figure \ref{figure:T_68_68V0} shows results obtained using the
reference potentials of equations (\ref{eqn:refunc2}) and
(\ref{eqn:refunc3}), again for $r_\text{match}=6.8$~\AA. The $C_6+C_8$
reference potential gives a marked improvement over the pure $C_6$
reference potential. The $\bm{T}$ matrix elements it produces follow
the form of the full coupled-channel results but still become poor at
energies much below 1~K: at $r_\text{match}=6.8$~\AA, $|
U_i^{\text{ref},C_6+C_8} - W_{ii}|/k_{\rm B}\approx 0.35$~K. However,
the results obtained with the $V_0$ reference potential are much more
accurate, and can scarcely be distinguished from the full
coupled-channel results in Figure \ref{figure:T_68_68V0}.

\begin{figure}[tb]
\centering
\includegraphics[width=1.0\columnwidth,clip]{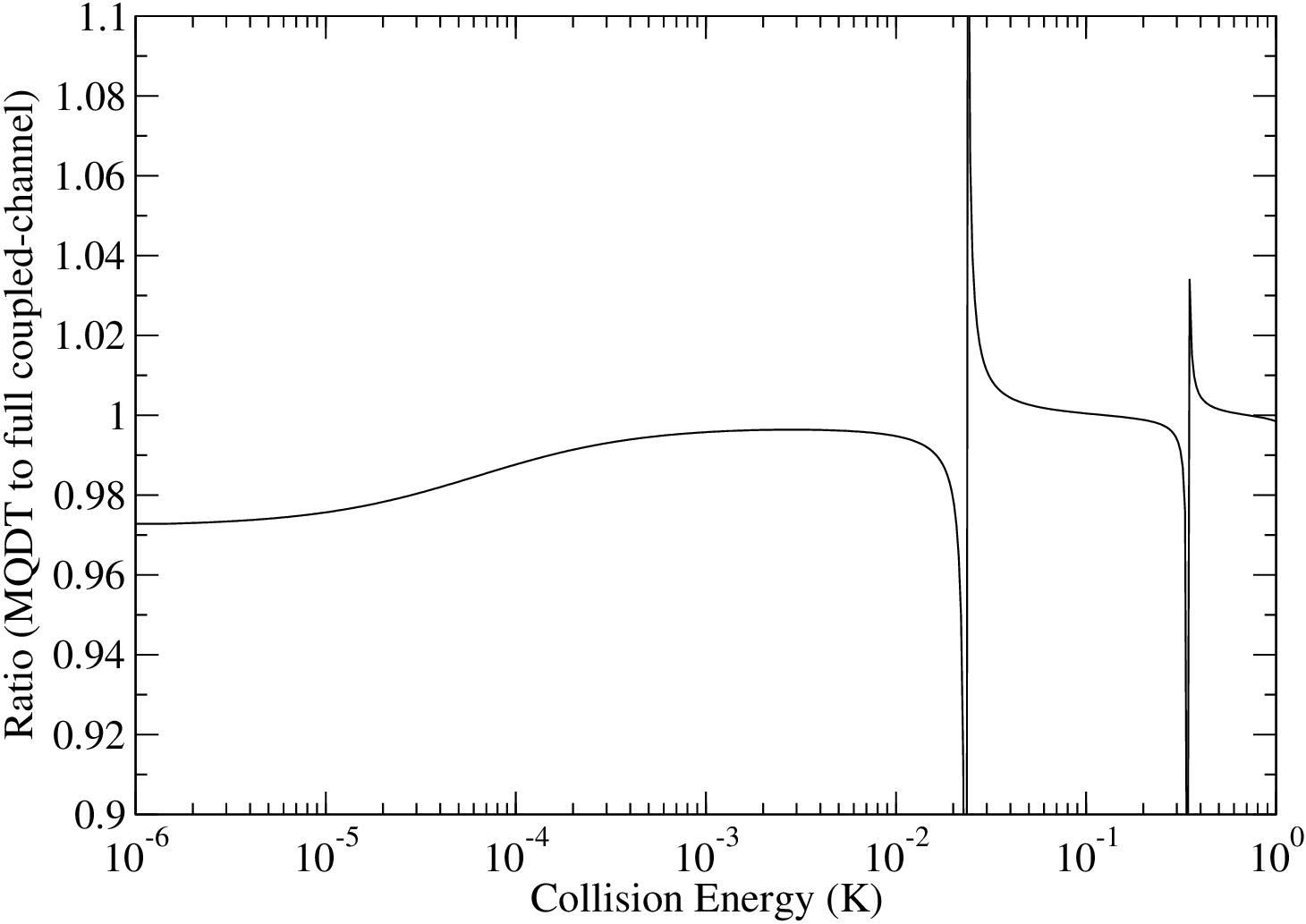}
\caption{
Ratio of the square of the diagonal $\bm{T}$ matrix element for
$m_j=+1$, $L=0$ at $B=10$~G for MQDT, with the $V_0$ reference
potential and $r_\text{match} = 6.8$~\AA, to that from full
coupled-channel calculations.} \label{figure:ratio_V0}
\end{figure}

Even the $V_0$ reference potential does not produce exact results.
Figure \ref{figure:ratio_V0} shows the ratio of the MQDT $T$-matrix
elements for this reference potential to the full coupled-channel
results. The poles in the ratio arise simply because MQDT places the
zeroes in $|T|^2$ (where the phase shift is an integer multiple of
$\pi$) at very slightly different collision energies. However, at very
low energies (below about 1~mK) the MQDT results underestimate the
squared $T$-matrix elements by up to 3\%. This probably arises because
the ``best" reference potential would be one that takes account of
adiabatic shifts due to mixing in excited rotational levels. For the
$n=0$ channels, the shift due to $n=1$ channels may be estimated from
2nd-order perturbation theory to be about 0.012 cm$^{-1}$ (equivalent
to 17~mK) at $r_\text{match} = 6.8$~\AA. This will cause residual
errors in the MQDT $C$ functions that are responsible for the small
errors visible in Figure \ref{figure:ratio_V0}.

\begin{figure}[tb]
\centering
\includegraphics[width=1.0\columnwidth,clip]{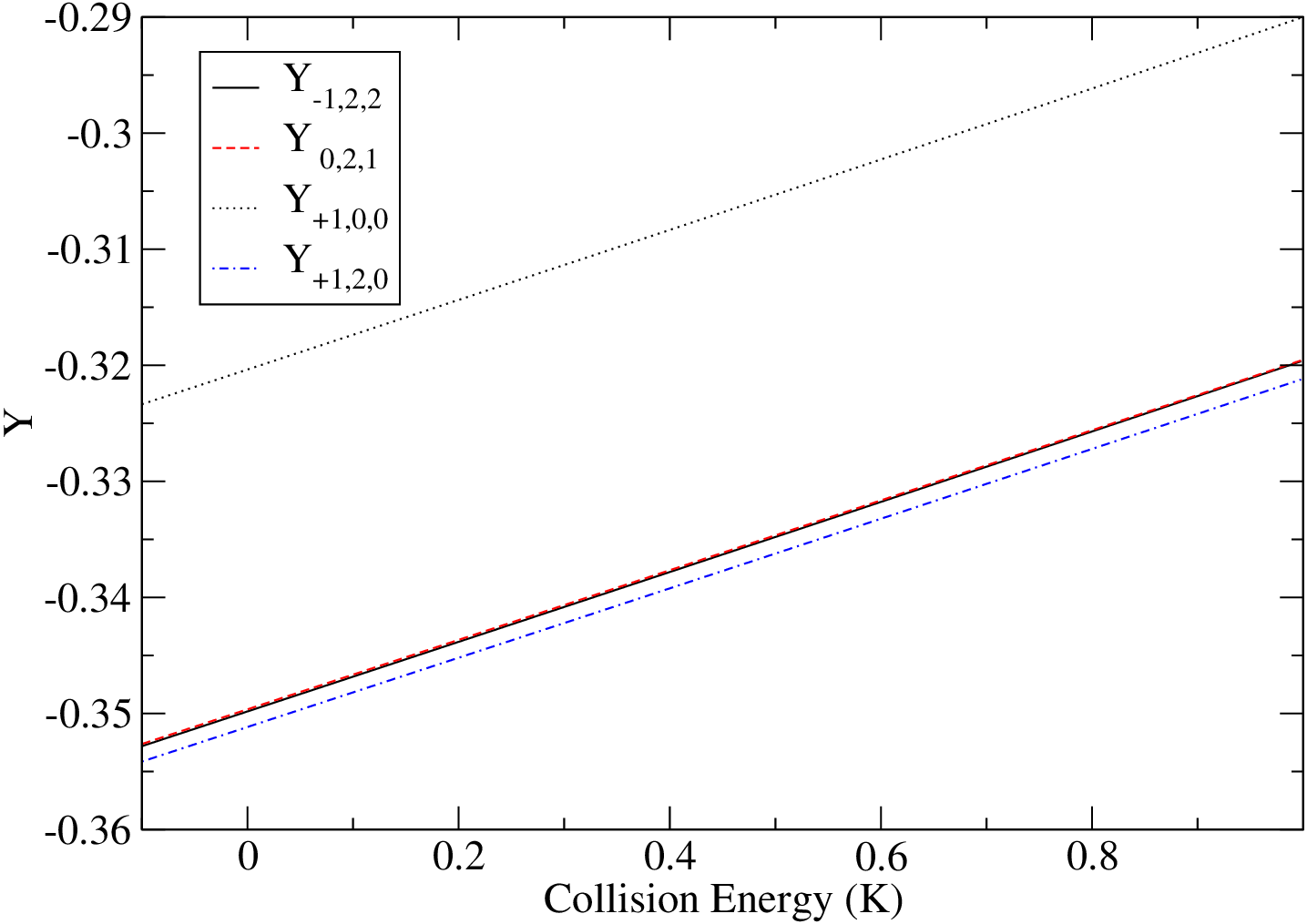}
\caption{(Color online)
Diagonal $\bm{Y}$ matrix elements as a function of energy at $B=10$~G,
for the $V_0$ reference potential with $r_\text{match} = 6.8$~\AA.}
\label{figure:Y_V0_68}
% T.eps: 0x0 pixel, 300dpi, 0.00x0.00 cm, bb=(atend)
\end{figure}

Figure \ref{figure:Y_V0_68} shows representative matrix elements of
$\bm{Y}$ obtained at $r_\text{match}=6.8$~\AA, with the $V_0$ reference
potential, as a function of energy. It may be seen that they are nearly
linear in energy. The other matrix elements of $\bm{Y}$ show similar
behavior. While the actual values of matrix elements vary
substantially, they are all nearly linear in energy for $r_\text{match}
= 6.8 \text{ \AA}$.

It should be noted that when the reference functions are obtained
numerically, as in the present work, there is no significant difference
in computer time for different choices of reference potential. Using
the full $V_0$ reference potential is just as inexpensive as using a
simpler one.

\subsubsection{Feshbach resonances}

Magnetic fields have important effects on cold molecular collisions,
and in particular magnetically tunable low-energy Feshbach resonances
provide mechanisms by which the collisions may be {\em controlled}. It
is therefore important to establish whether the $\bm{Y}$ matrices
obtained from MQDT are smooth functions of magnetic field as well as
energy and can be used to characterize Feshbach resonances. If they
are, it will offer substantial computational efficiencies.

\begin{figure}[tb]
\centering
\includegraphics[width=1.0\columnwidth,clip]{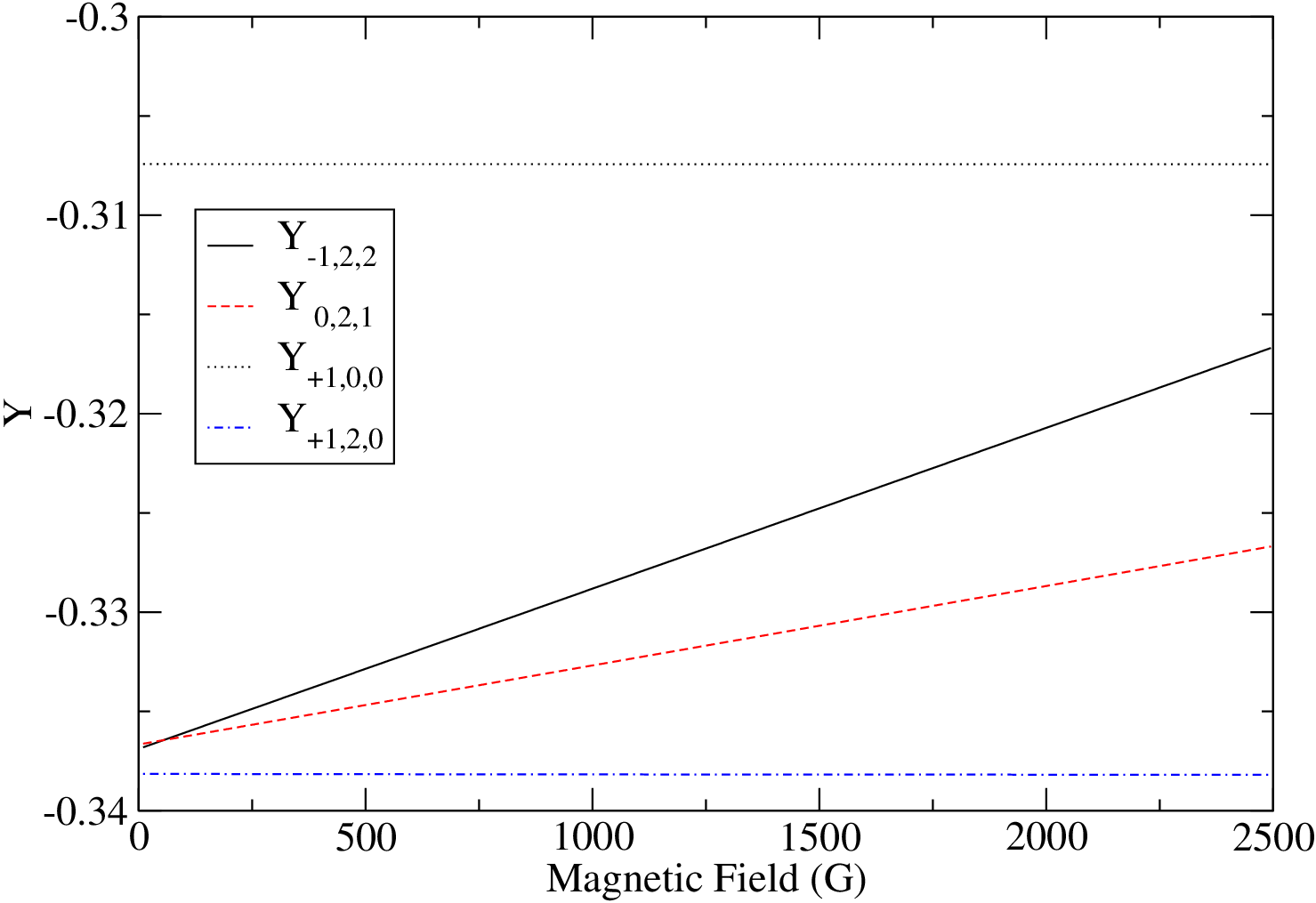}
\caption{(Color online)
Diagonal $\bm{Y}$ matrix elements as a function of magnetic field at
$E/k_{\rm B}=400$~mK, for the $V_0$ reference potential with
$r_\text{match} = 6.8$~\AA.} \label{figure:Y_400}
% T.eps: 0x0 pixel, 300dpi, 0.00x0.00 cm, bb=(atend)
\end{figure}

\begin{figure}[tb]
\centering
\includegraphics[width=1.0\columnwidth,clip]{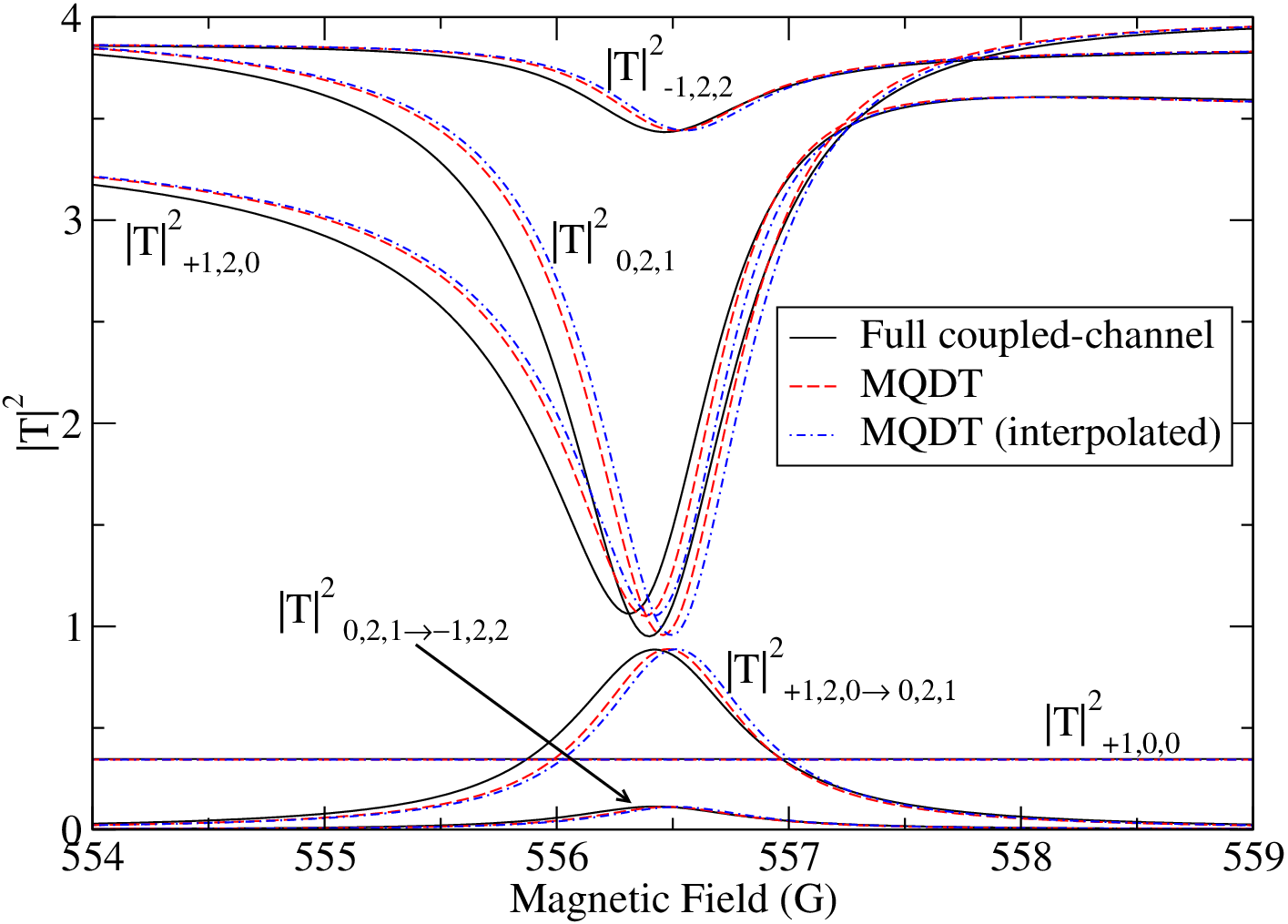}
\includegraphics[width=1.0\columnwidth,clip]{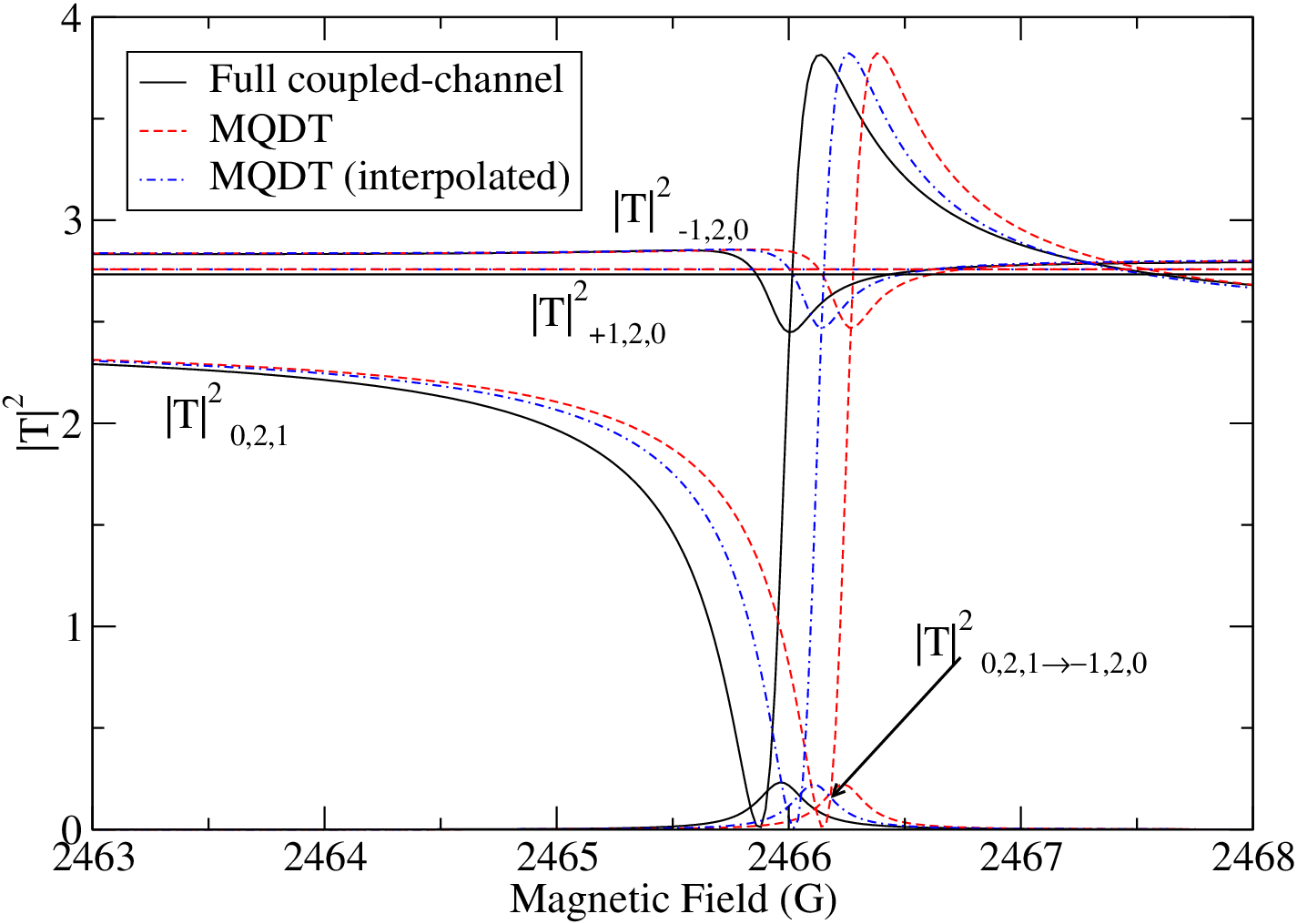}
\caption{(Color online) Squares of diagonal and off-diagonal $T$-matrix
elements as the field is tuned across a Feshbach resonance at $E/k_{\rm
B}=400$~mK (upper panel) and 1~mK (lower panel). The MQDT results are
obtained with the $V_0$ reference potential at $r_\text{match} =
6.8$~\AA.} \label{figure:T_FESH}
% T.eps: 0x0 pixel, 300dpi, 0.00x0.00 cm, bb=(atend)
\end{figure}

Figure \ref{figure:Y_400} shows how the diagonal $\bm{Y}$ matrix
elements vary as a function of magnetic field for Mg+NH collisions over
the range from 0 to 2500~G for a collision energy of 400~mK. It may be
seen that the matrix elements are indeed very nearly linear, as
required for efficient interpolation.

In Mg+NH, there is a Feshbach resonance due to the $n=1$, $j=0$,
$m_j=0$, $L=3$ state shown in Figure \ref{figure:Y_C6_20} that tunes
down towards the $n=0$, $m_j=+1$ threshold with increasing field.
Figure \ref{figure:T_FESH} shows the comparison between MQDT and full
coupled-channel calculations for a selection of diagonal and
off-diagonal $\bm{T}$ matrix elements as the magnetic field is tuned
across this resonance at energies of 400~mK and 1~mK. At each energy,
MQDT results were obtained both by recalculating the $\bm{Y}$ matrix at
every field and by linear interpolation between two points separated by
100~G. In both cases, the interpolated MQDT results are within about
0.2~G of the full MQDT results even for this long interpolation, and
this could of course be improved simply by considering a few more
fields across the range to allow. However, there is also a residual
error of of 0.1 to 0.2~G in the resonance position even for the full
MQDT results, which is not very different at the two collision energies
considered. This is again likely to be due to the effect as described
in Section \ref{sec:refchoice}: the $V_0$ reference potential neglects
couplings between channels outside $r_\text{match}$, and for the small
value of $r_\text{match}$ used here these couplings are sufficient to
shift the resonance positions slightly. Apart from these small shifts,
however, both the elastic and the inelastic scattering around the
resonances are very well described at both energies.

The linearity of the $\bm{Y}$ matrix with both energy and applied
magnetic field is an extremely promising result, and suggests that MQDT
will provide very efficient ways of performing cold collision
calculations as a function of energy and magnetic field, without
needing to repeat the expensive coupled-channel part of the calculation
on a fine grid.

\section{Conclusions}

We have shown that Multichannel Quantum Defect Theory (MQDT) can be
applied to low-energy molecular collisions in applied magnetic fields.
MQDT provides a matrix $\bm{Y}$, defined at a distance $r_\text{match}$
at relatively short range, which encapsulates all the short-range
dynamics of the system. For the prototype Mg+NH system, we have shown
that MQDT can provide numerical results that are in quantitative
agreement with full coupled-channel calculations if the MQDT reference
functions are defined appropriately.

We have investigated the effect of different choices of reference
potential and values of $r_\text{match}$. For cold atom-molecule
collisions, unlike cold atom-atom collisions, calculations are likely
to be needed over a significant range of collision energy, perhaps 1~K
or so. If $r_\text{match}$ is placed at too long a range, there is a
significant likelihood of resonant features within the energy range
that prevent simple interpolation of $\bm{Y}$. This may be circumvented
by carrying out the matching at a smaller distance $r_\text{match}$.
However, when this is done, a pure $C_6$ reference potential may not be
sufficient. For Mg+NH, the most satisfactory procedure is to perform
matching at fairly short range (inside 7~\AA) and use a reference
potential that is defined to be the same as the true diagonal potential
in the incoming channel.

The major strength of MQDT for molecular applications is that, if the
the matching to obtain $\bm{Y}$ is carried out at relatively short
range, the matrix is only weakly dependent on collision energy and
magnetic field. This allows very considerable computational
efficiencies, because the expensive calculation to obtain $\bm{Y}$
needs to be carried out at only one or a few combinations of collision
energy and field. The remaining calculations to obtain scattering
properties on a fine grid of energies and fields are then
computationally inexpensive, varying only linearly with the number of
channels $N$. Full coupled-channel calculations, by contrast, scale as
$N^3$.

MQDT is a promising alternative to full coupled-channel calculations
for cold atom-molecule collisions, particularly when fine scans over
collision energy and magnetic field are required. In future work, we
will investigate further the choice of reference functions to optimize
the accuracy and to minimize the dependence of $\bm{Y}$ on collision
energy and field. We will also investigate how the results for Mg+NH
transfer to more strongly anisotropic systems, with stronger long-range
anisotropy and more closed channels that are capable of producing
scattering resonances.

\section{Acknowledgments}

JMH is grateful to Chris Greene and John Bohn for interesting him in
this project. The authors thank Maykel Leonardo Gonz\'alez-Mart\'inez
and Jesus Aldegunde for exploratory work on molecular applications of
MQDT at the beginning of the project. JFEC is grateful to EPSRC for a
High-End Computing Studentship.

\bibliography{../../all}

\end{document}